\UseRawInputEncoding
\documentclass[twocolumn]{aastex63}
\usepackage{amsmath,mathtools,mathrsfs}
\usepackage[T1]{fontenc}

\begin{document}
\title{Brightness Asymmetry of Black Hole Images as a Probe of Observer Inclination}

\author{Lia Medeiros}
\altaffiliation{NSF Astronomy and Astrophysics Postdoctoral Fellow}
\altaffiliation{email: lia@ias.edu}
\affiliation{School of Natural Sciences, Institute for Advanced Study, 1 Einstein Drive, Princeton, NJ 08540}
\author{Chi-Kwan Chan}
\affiliation{Steward Observatory and Department of Astronomy, University of Arizona, 933 N. Cherry Ave., Tucson, AZ 85721}
\affiliation{Data Science Institute, University of Arizona, 1230 N. Cherry Ave., Tucson, AZ 85721}
\affiliation{Program in Applied Mathematics, University of Arizona, 617 N. Santa Rita, Tucson, AZ 85721}
\author{Ramesh Narayan}
\affiliation{Center for Astrophysics $|$ Harvard \& Smithsonian, 60 Garden Street, Cambridge, MA 02138, USA}
\author{Feryal \"Ozel}
\affiliation{Steward Observatory and Department of Astronomy, University of Arizona, 933 N. Cherry Ave., Tucson, AZ 85721}
\author{Dimitrios Psaltis}
\affiliation{Steward Observatory and Department of Astronomy, University of Arizona, 933 N. Cherry Ave., Tucson, AZ 85721}

\begin{abstract}
The Event Horizon Telescope recently captured images of the supermassive black hole in the center of the M87 galaxy, which show a ring-like emission structure with the South side only slightly brighter than the North side. This relatively weak asymmetry in the brightness profile along the ring has been interpreted as a consequence of the low inclination of the observer (around $17^\circ$ for M87), which suppresses the Doppler beaming and boosting effects that might otherwise be expected due to the nearly relativistic velocities of the orbiting plasma. In this work, we use a large suite of general relativistic magnetohydrodynamic simulations to reassess the validity of this argument. By constructing explicit counter examples, we show that low-inclination is a sufficient but not necessary condition for images to have low brightness asymmetry. Accretion flow models with high accumulated magnetic flux close to the black hole horizon (the so-called magnetically arrested disks) and low black-hole spins have angular velocities that are substantially smaller than the orbital velocities of test particles at the same location. As a result, such models can produce images with low brightness asymmetry even when viewed edge on.
\end{abstract}

 
%==============================================================================
\section{Introduction}\label{sec:intro}

The first images of a supermassive black hole at event horizon scale resolution \citep{2019ApJ...875L...1E, 2019ApJ...875L...2E, 2019ApJ...875L...3E, 2019ApJ...875L...4E, 2019ApJ...875L...5E, 2019ApJ...875L...6E,2021ApJ...910L..12E,2021ApJ...910L..13E} have opened up a new avenue for studying the characteristics of black holes and their accretion flows. The two main targets for the Event Horizon Telescope (EHT), i.e., the black holes in the center of the M87 galaxy and of our own galaxy (Sagittarius A*, hereafter Sgr A*), are both fed by low-luminosity accretion flows. In this regime, their spectral and imaging properties are best explained in the context of geometrically thick, advection-dominated accretion flows (ADAFs, also called radiatively inefficient accretion flows, RIAFs; see \citealt{2014ARA&A..52..529Y} for a review). As predicted theoretically~\citep{2000ApJ...541..234O} and demonstrated with the EHT images, these flows are optically thin at millimeter wavelengths and, therefore, transparent to synchrotron emission down to the black-hole horizon. The millimeter images of these black holes are characterized by a bright ring of emission that surrounds a deep brightness depression: the shadow of the black hole.

The outline of the black hole shadow encodes signatures of the black hole spacetime (see, e.g., \citealt{2010ApJ...718..446J,2020ApJ...896....7M,2020PhRvL.125n1104P}). On the other hand, the azimuthal brightness profile of the emission ring carries information primarily about the velocity structure in the accretion flow itself. Relativistic Doppler beaming and boosting causes the region of the accretion flow with velocities directed towards the observer to appear brighter in the image and the receding side to appear dimmer. As a result, sources that are viewed edge-on are expected to have large Doppler asymmetries and, therefore, crescent image shapes; sources that are viewed face-on result in a more symmetric ring-link emission structure, as is the case of the M87 image\footnote{Hereafter we will refer to differences in the brightness of the ring as a function of position angle along the ring as image brightness asymmetry to distinguish it from asymmetry in the shape of the ring such as an overall deviation from circularity.}. Such arguments have been used to motivate the use of image brightness asymmetry to measure the inclination of the observer's line of sight with respect to the angular momentum axis of the accretion flow -- and also to the spin axis of the black hole, if the latter is assumed to be aligned with the former \citep{2015ApJ...798...15P}. 

\begin{figure}[t!]
\centering
\includegraphics[width=\columnwidth]{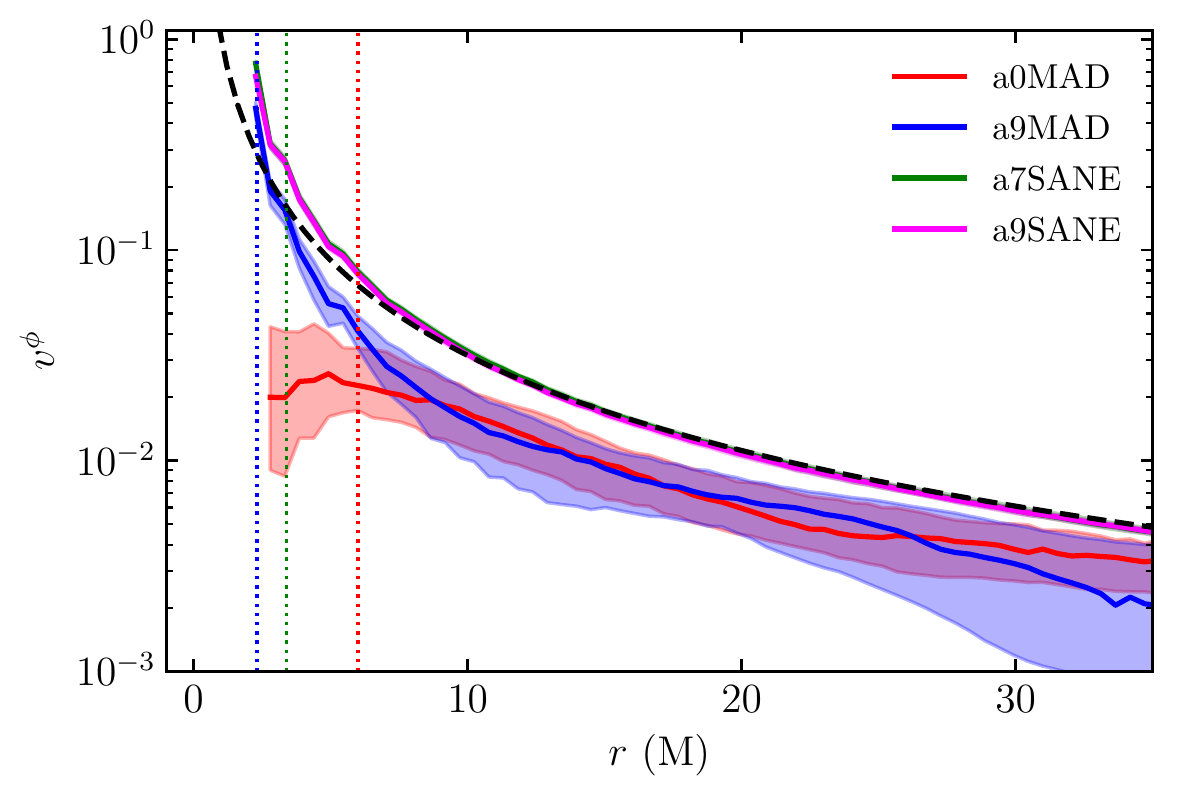}
\caption{The azimuth- and time-averaged angular velocity ($v_{\phi}$) as a function of radius for the four GRMHD simulations we consider here (the curves for the two SANE simulations are overlapping). The curves show the median while the shaded regions show the $25\%$ and $75\%$ bounds. The black dashed curve corresponds to the Keplerian profile ($v^{\phi}=r^{-3/2}$). The dotted vertical lines correspond to the location of the ISCO for a black hole with $a_{{\rm BH}}=0.0$ (red), $a_{{\rm BH}}=0.7$ (green), and $a_{{\rm BH}}=0.9$ (blue).}
\label{fig:vphi}
\end{figure}

In this paper, we re-evaluate the above argument and explore whether it is possible to create an image with low image brightness asymmetry without requiring a low inclination for the observer. Because the asymmetry is caused by relativistic Doppler beaming, a lower angular velocity for the emitting matter could result in significantly lower image brightness asymmetry even at high observer inclinations. Angular velocities that are low compared to the orbital velocities of test particles are frequently seen in the so-called Magnetically Arrested Disks (MAD, see e.g., \citealt{2012MNRAS.426.3241N}), where magnetic forces close to the black hole alter significantly the angular momentum of the infalling plasma  (see also \citealt{2021MNRAS.501.4722B} for a brief discussion of the lower angular velocities in MAD models). This is contrary to the case of the Standard and Normal Evolution models (SANE, see e.g., \citealt{2003ApJ...592.1042I}) or the semi-analytic models used in \citet{2015ApJ...798...15P} (see Figure~\ref{fig:vphi} below and Figures~12 and 20 in \citealt{2012MNRAS.426.3241N} for a comparison of the angular velocity profiles of MAD and SANE flows). We use the results of detailed General Relativistic MagnetoHydrodynamic (GRMHD) simulations to show that, indeed, low inclination is not necessary to create images with low image brightness asymmetry.  The low plasma velocities near the horizons of low-spin MAD models can create relatively symmetric images even at high observer inclinations. These MAD models are particularly relevant to M87, as they have been shown to be in better agreement with recent EHT polarization results compared to SANE models \citep{2021ApJ...910L..12E,2021ApJ...910L..13E}. Even though we specifically discuss the implications of our results for the image of the black hole in M87, the image symmetry considerations are more general and, therefore, more broadly applicable.

This paper is organized as follows. In Section~\ref{sec:sims}, we discuss the GRMHD simulation library used in this work. Section~\ref{sec:Doppler} discusses how relativistic Doppler effects affect image brightness asymmetry. In Section~\ref{sec:asym_def}, we introduce a definition for image brightness asymmetry, and in Section~\ref{sec:results}, we explore how image brightness asymmetry correlates with black hole and accretion flow parameters. We discuss the implications of our work in Section~\ref{sec:discussion} and summarize our findings in Section~\ref{sec:summary}.

%==============================================================================
\section{GRMHD+radiative transfer simulations}\label{sec:sims}

We employ a set of GRMHD simulations that were performed using the {\tt HARM3D} code \citep{2003ApJ...589..444G} and that were first discussed in \citet{2012MNRAS.426.3241N} and \citet{2013MNRAS.436.3856S}. In each simulation, the accretion flow was evolved from a torus located between $r_{\rm in, \, sim}=10M$ and $r_{\rm out, \, sim}=1000M$ with a peak density around $r_{\mathrm{max}}\approx 20M$, and the flow had an adiabatic index of $\gamma=5/3$. The simulations were run for a long time span, $t=200,000~GM/c^3$, such that steady state conditions were reached in the inner flows. The set of simulations includes a non-spinning MAD model (hereafter a0MAD), a MAD model with $a_{\mathrm{BH}}=0.9$ (hereafter a9MAD), a SANE model with $a_{\mathrm{BH}}=0.7$ (hereafter a7SANE) and a SANE model with $a_{\mathrm{BH}}=0.9$ (hereafter a9SANE). For more details on the properties of these simulations, see \citet{2012MNRAS.426.3241N, 2013MNRAS.436.3856S}. 

%%%%%%%%%%%%%%%%%%%%%%%%%%%%%%%%%

We perform new radiative transfer and ray-tracing simulations using the fast GPU-based code {\tt GRay} \citep{2013ApJ...777...13C}. Non-radiative GRMHD simulations are invariant to a rescaling of the number density of the plasma particles but the radiative transfer calculations are not. For this reason, we consider five different values for the number density scale of electrons, for each of these four GRMHD simulations: $n_{e,0}=10^6 \; \mathrm{cm}^{-3}, \, 5\times 10^6 \; \mathrm{cm}^{-3}, \, 10^7 \; \mathrm{cm}^{-3}, \, 5\times 10^7 \; \mathrm{cm}^{-3},$ and $ 10^8\; \mathrm{cm}^{-3}$, where $n_{e,0}$ corresponds approximately to the number density at horizon scales. We also consider four values for the observer inclination ($i=0^{\circ}, \, 19^{\circ}, \, 42^{\circ},\, 90^{\circ}$), which are evenly spaced in $\sin i$ given that Doppler effects scale as $v\sin i$. 

GRMHD simulations evolve only the internal energy density of the plasma, which allows us to calculate only the temperature of the ions and not of the electrons. We, therefore, employ a temperature prescription for the electrons that sets the ion-to-electron temperature ratio $T_{\rm i}/T_{\rm e}$ based on the local value of the parameter $\beta=P_{\mathrm{gas}}/P_{\mathrm{mag}}$ \citep{2015ApJ...799....1C}, defined as the ratio of gas pressure to magnetic field pressure, and is given by \citep{2016A&A...586A..38M, 2019ApJ...875L...4E}
\begin{equation}
\frac{T_i}{T_e}=R_{\mathrm{high}}\frac{\beta^2}{1+\beta^2} +\frac{1}{1+\beta^2}.
\end{equation}
We explore three values for the $R_{\mathrm{high}}$ parameter ($R_{\mathrm{high}}=1,\, 20, \, 80$)\footnote{Note that a $R_{\mathrm{high}}$ value of unity would set the electron temperature equal to the ion temperature, which would result in a model that is unrealistic for the two low-luminosity sources the EHT can resolve. We include these models for consistency with previous work but use $R_{\mathrm{high}}=20$ as our default value.} in our suite of simulations.

%%%%%%%%%%%%%%%%%%%%%%%%%%%%%%%%%
\begin{figure*}[t!]
\centering
\includegraphics[width=.9\textwidth]{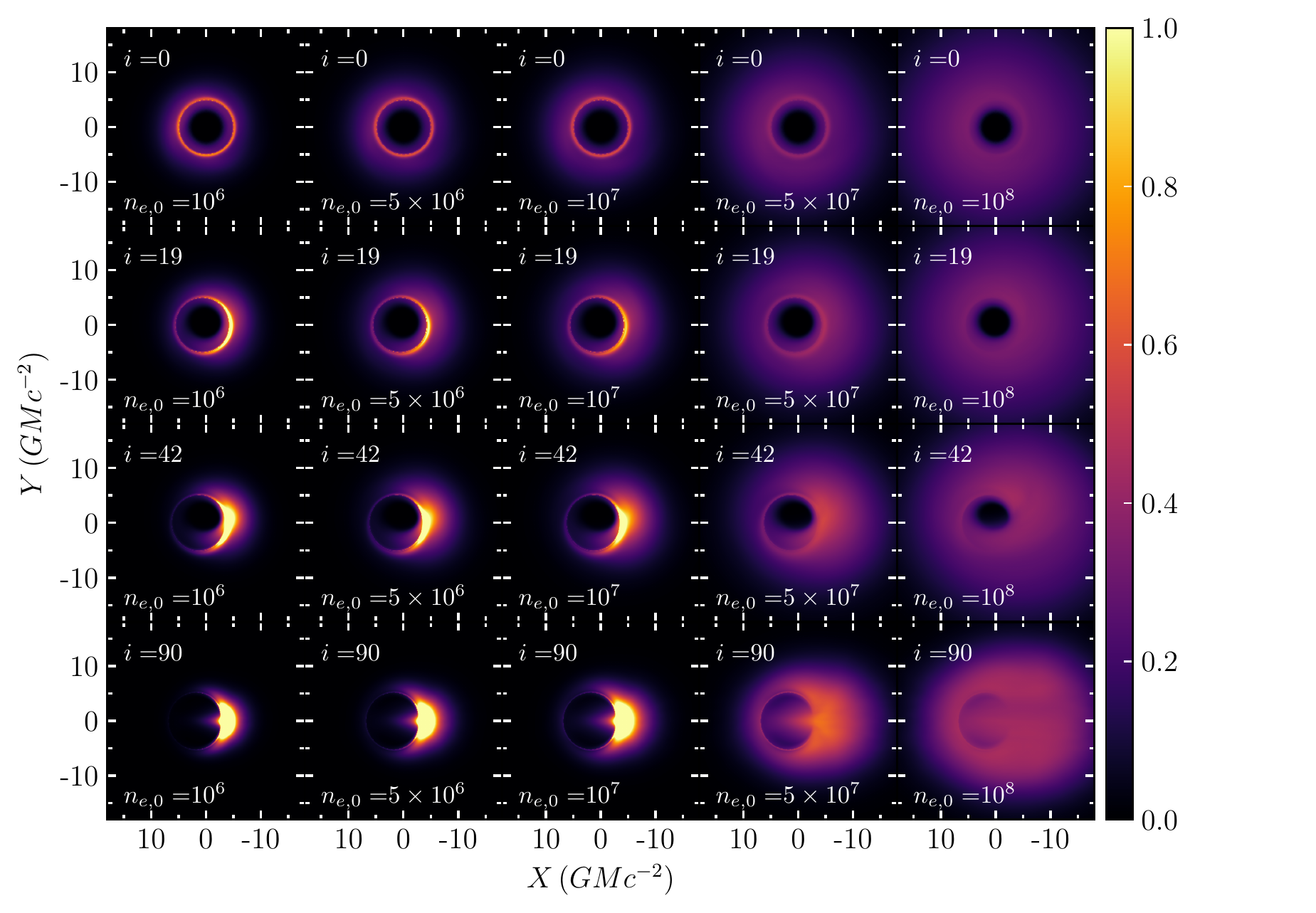}
\caption{Effect of changing the electron density scale $n_{e,0}$ (different columns) and the observer inclination $i$ (different rows) on the average image of a SANE simulation with $a_{\rm BH}=0.9$ spin and $R_{\mathrm{high}}=20$. In all panels the orientation angle of the spin axis on the plane of the sky ($\psi$) is set to zero so that the black-hole spin points upward. Each panel has been normalized such that all panels with the same value of $n_{e,0}$ have the same total flux. The intensity values in the colorbar are in arbitrary units. Here and in the following figures, the images were calculated at a wavelength of 1.3mm and the snapshots were averaged over a timespan of $10,240GMc^{-3}$ with a resolution of $10GMc^{-3}$.}
\label{fig:ne_i9SANE}
\end{figure*}

For the radiative transfer calculations, we further assume a black hole mass of $4.25\times10^6M_{\odot}$ and a distance of $8.3\mathrm{kpc}$. These values were chosen to be consistent with Sgr~A* and result in a mass to distance ratio that is comparable to the values measured by the GRAVITY collaboration and the UCLA galactic center group (see e.g., \citealt{2019Sci...365..664D,2020A&A...636L...5G,2021A&A...647A..59G}). However, this choice does not limit the generality of our results, because the emissivity of the accretion flow at 1.3\;mm approximately depends on the quantity $M_{\mathrm{BH}}n_{e,0}^2$, making the black hole mass and the electron number density scale degenerate (see Appendix A of Satapathy et al. 2021, in prep; see also \citealt{2015ApJ...799....1C}). Indeed, when we compare directly our simulations to the M87 results in Section~5, we use a black hole mass of $6.5\times10^9M_{\odot}$ \citep{2019ApJ...875L...1E} and shift the range of $n_{e,0}$ values accordingly.

%%%%%%%%%%%%%%%%%%%%%%%%%%%%%%%%%
\begin{figure*}[!htb]
\centering
\includegraphics[width=.9\textwidth]{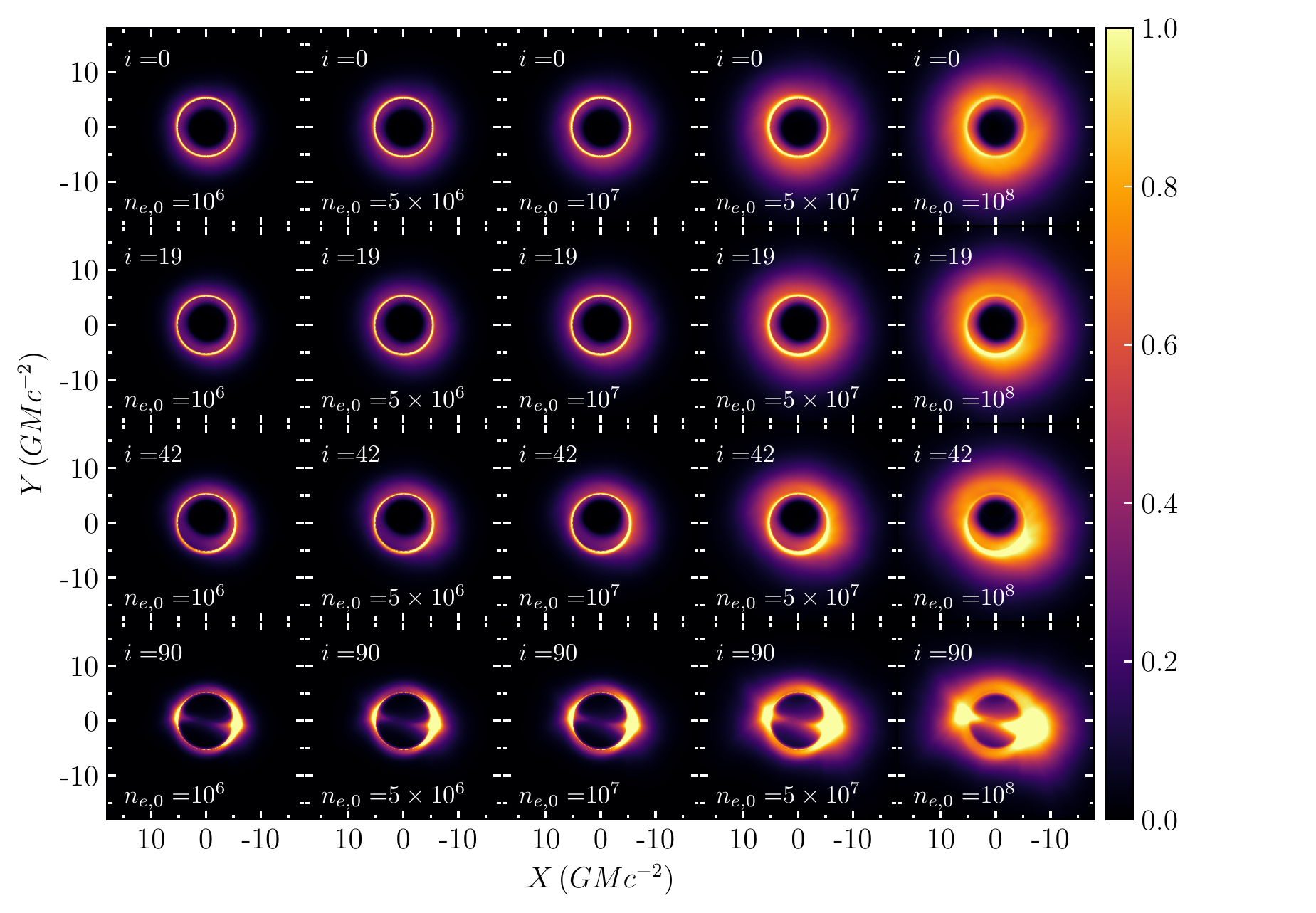}
\caption{Same as Figure \ref{fig:ne_i9SANE} but for a MAD simulation with $a_{\mathrm{BH}}=0$ spin and $R_{\mathrm{high}}=20$.}
\label{fig:ne_i0MAD}
\end{figure*}

%%%%%%%%%%%%%%%%%%%%%%%%%%%%%%%%%
\begin{figure*}[!htb]
\centering
\includegraphics[width=.78\textwidth]{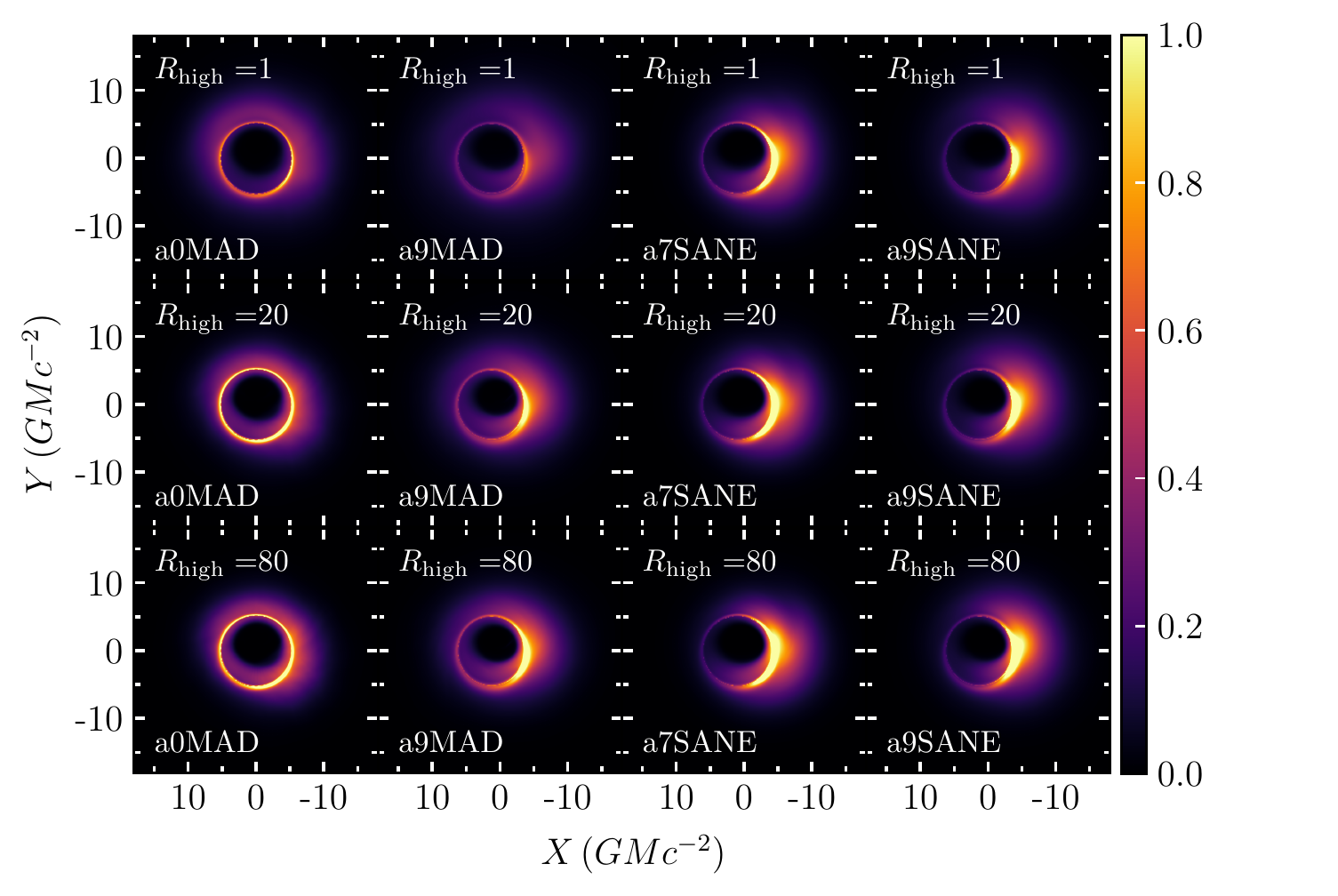}
\caption{Similar to Figure \ref{fig:ne_i0MAD} but showing the effect of changing the ion-to-electron temperature ratio $R_{\mathrm{high}}$ (different rows) for four GRMHD simulations. For all models in this figure we have set the electron number density scale to $n_e=10^7\mathrm{cm}^{-3}$, the inclination to $i=42^{\circ}$, and all panels have been normalized such that they have the same total flux.}
\label{fig:GRMHD_R}
\end{figure*}
%\newpage

The parameter exploration discussed above resulted in 240 simulations, each with 1024 snapshots with $10GM/c^3$ temporal resolution, a field of view of $64GM/c^2$, and a spatial resolution of $1/8M$ (see \citealt{2020arXiv200406210P} for an exploration of optimal pixel resolution). The SANE simulation with spin 0.7 and observer inclination $i=0^{\circ}$ was removed from our simulation library due to numerical artifacts caused by the pole of the Boyer-Lindquist coordinate system when viewed at such low inclination. This leaves a total of 225 simulations with 1024 snapshots each for a total of $230,400$ image snapshots.

%==============================================================================
\section{Relativistic Doppler Effects in Black-Hole Images}\label{sec:Doppler}

In Figure~\ref{fig:vphi} we show the azimuth- and time-averaged angular velocities ($v_{\phi}$) for the four GRMHD simulations we consider. The two SANE simulations follow the Keplerian profile far from the black hole (as expected) and exceed it slightly close to the black hole. The non-spinning MAD simulation, however, is significantly below the Keplerian profile and the profile flattens out close to the black hole due to the high magnetic flux. The MAD simulation with spin $a_{\mathrm{BH}}=0.9$ is also significantly below the Keplerian profile far from the black hole but exceeds the Keplerian velocity close to the black hole. Because our simulation library includes a limited sample of spins, we leave a detailed exploration of the dependence of image brightness asymmetry on spin to future work.

Figure~\ref{fig:ne_i9SANE} shows the effect of changing the electron number density scale and the observer inclination on the average image of the simulations for a representative SANE GRMHD model. Because in this simulation the plasma velocity in the inner accretion flow is comparable to the near-relativistic orbital velocity of test particles, increasing the observer inclination results in a high degree of left-right asymmetry (relative to the angular momentum vector that points upward) due to the Doppler beaming and boosting effects. The plasma on the right side of the shadow moves with very high velocity towards us, causing that side of the image to be substantially brighter than the left, receding side.

The situation is markedly different in the $a_{\mathrm{BH}}=0$ MAD GRMHD simulation shown in Figure~\ref{fig:ne_i0MAD}, where the left-right brightness asymmetry of the images remains marginal, even at high observer inclinations. This is, of course, a direct consequence of the fact that magnetic stresses at the inner accretion flow substantially reduce the magnitudes of the orbital velocities of the plasmas and, hence, of the Doppler asymmetry of the images. 

As expected, increasing the electron number density scale increases the width of the ring in the images. As the ring of emission becomes wider, a broader range of annuli in the accretion flow contribute to the image brightness. Originating at larger distances from the horizon, from plasmas with smaller velocities, the photons emerging from such annuli experience a smaller degree of gravitational lensing and Doppler boosting and the brightness asymmetry of the image is dictated more by the geometric thickness of the flow and projection effects rather than by Doppler effects. 

Finally, in Figure \ref{fig:GRMHD_R} we show the effect of changing the ratio of the ion to the electron temperature $R_{\mathrm{high}}$ on the average image of the four simulations for a particular choice of electron number density scale and inclination. Changing $R_{\mathrm{high}}$ only has a marginal effect on the image brightness asymmetry, as we will quantify in more detail in the next section.

%%%%%%%%%%%%%%%%%%%%%%%%%%%%%%%%%
\begin{figure*}[t!]
\centering
\includegraphics[height=.75\columnwidth]{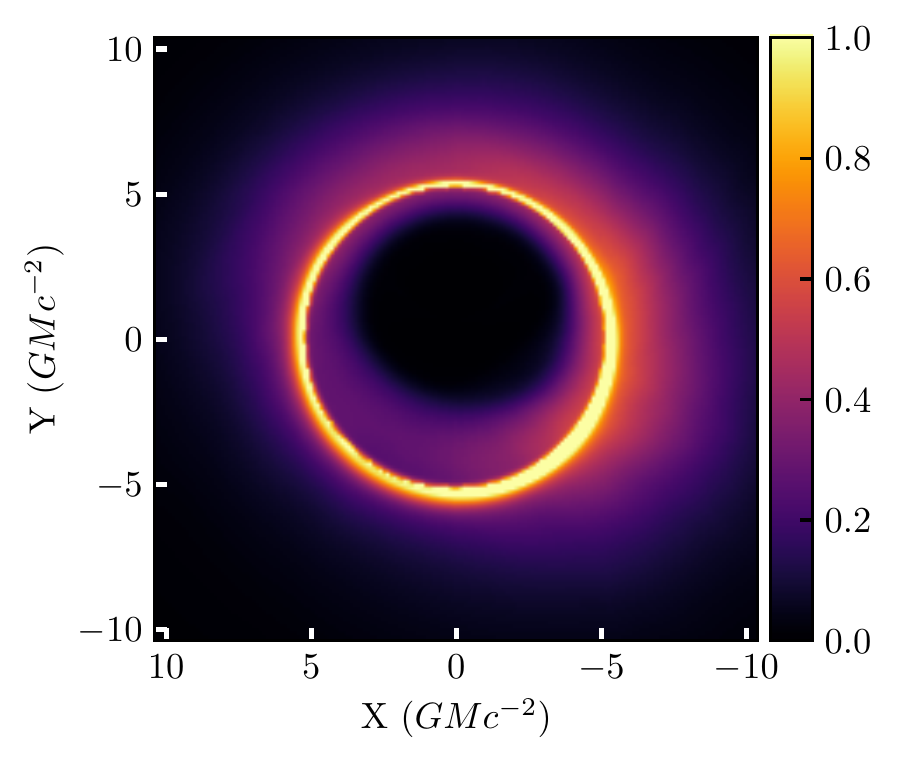}
\includegraphics[height=.75\columnwidth]{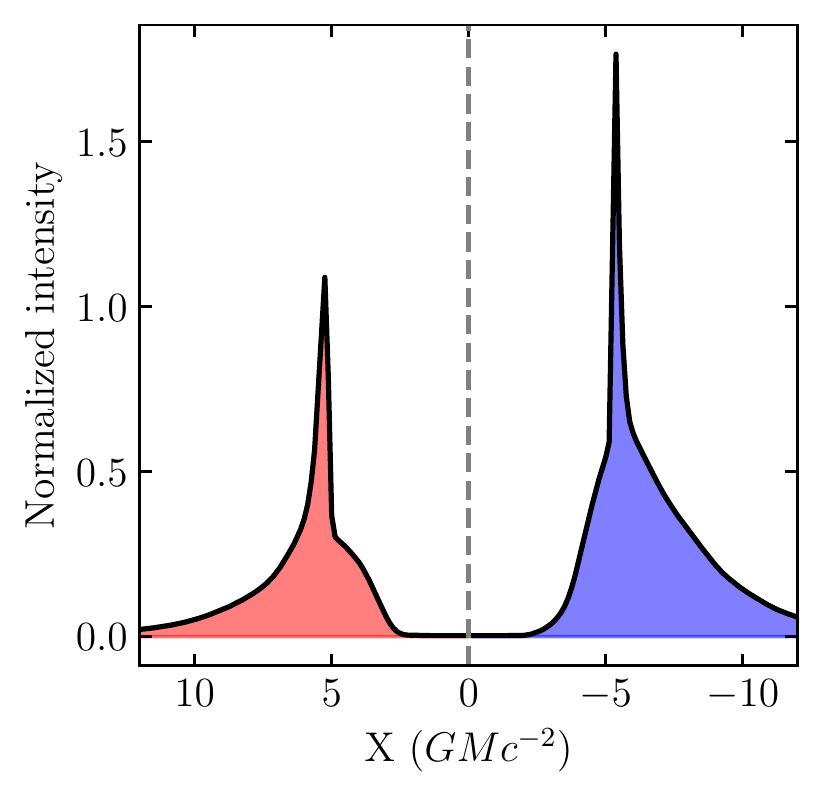}
\caption{({\it left}) The average image of a non-spinning MAD simulation with $i=42^{\circ}$, $R_{\mathrm{high}}=20$, and $n_{e,0} = 10^7\mathrm{cm}^{-3}$. ({\it right}) Horizontal cross section of the average image shown on the left panel, split in two halves. We define the image brightness asymmetry $A$ as the ratio of the half with the larger integral over $X$ (the shaded blue half in this case) over the half with the smallest integral (the shaded red half in this case). Because of the brightness depression in the center of the image, the exact location of the boundary between the two halves does not significantly affect our asymmetry measure.}
\label{fig:cross_sec}
\end{figure*}

%%%%%%%%%%%%%%%%%%%%%%%%%%%%%%%%%
\begin{figure}[t!]
\centering
\includegraphics[width=\columnwidth]{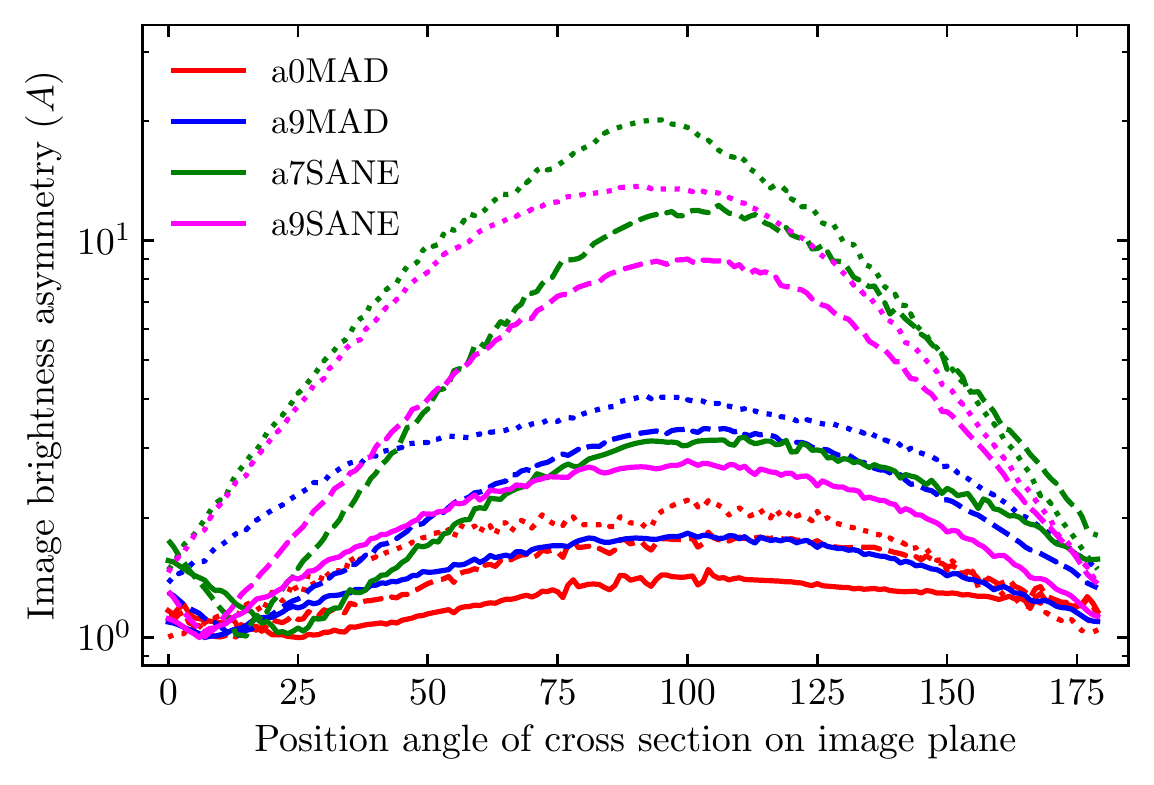}
\caption{Image brightness asymmetry $A$ calculated along different cross-sections on the image plane, at various position angles; a position angle of $90^{\circ}$ corresponds to a cross section which is perpendicular to the black hole spin axis. Different colors correspond to the four different GRMHD simulations (with $n_{e,0}=10^7$~cm$^{-3}$ and $R_{\mathrm{high}}=20$) and different line styles correspond to different observer inclination angles (solid: $i=19^\circ$; dashed: $i=42^\circ$; dotted: $i=90^\circ$). In all cases we have used the average images of the simulations. The peak image brightness asymmetry occurs for cross sections perpendicular to the black hole spin axis demonstrating that this asymmetry is caused primarily by relativistic Doppler effects.}
\label{fig:asym_angle}
\end{figure}

%==============================================================================
\section{Image Brightness Asymmetry}\label{sec:asym_def}

There are numerous approaches one could follow to quantify the brightness asymmetry of a ring-like image, such as decomposing it to polar harmonics or measuring the rms amplitude of its brightness along a circular path. Among the various definitions, we choose here, through trial and error, a particular one that emphasizes the dependence of this asymmetry on the various model parameters. 

We start by setting a Cartesian $(X,Y)$ coordinate system on the image plane with the $Y$-axis parallel to the spin angular momentum of the flow and the black hole, as in Figures~\ref{fig:ne_i9SANE} and \ref{fig:ne_i0MAD}. The center of the black-hole shadow and, hence, of the image is displaced because of the differential frame dragging effects by $2a_{\rm BH}\sin i$ along the $X$-direction (see, e.g., \citealt{2006PhRvD..74f3001B,2010ApJ...718..446J}). We expect the maximal brightness asymmetry caused by Doppler effects to occur along the $Y=0$ cross section of the image. For this reason, we consider the brightness of the image along this cross section, which we denote by $I(X,Y=0)$, and define the brightness asymmetry $A$ as the ratio between the brightness integrated over the two halves of this cross-section to the left and to the right of the image center. We further constrain this ratio to be greater than unity by setting the half of the cross section that has the greatest integral as the numerator. This yields
\begin{equation}
    A =  \frac{\int_{2 a\sin(i)}^{r_{\mathrm{out}}} I(X,Y=0) dX}{\int^{2a\sin(i)}_{-r_{\mathrm{out}}} I(X,Y=0) dX}\;,
\end{equation}
if this ratio is larger than unity, or the inverse of it if it is not. In this expression, we have  omitted the subscript from $a_{\mathrm{BH}}$ for brevity and set $r_{\mathrm{out}}$ to the outer radius of the simulated images, which is at $r_{\mathrm{out}}=32GMc^{-2}$. Equation (2) is defined such that it matches the asymmetry one would measure from a reconstructed black hole image. The definition of $A$ is insensitive to the precise boundary between the two sides of the image, which appears in the limits of the integrals in the numerator and the denominator,  because of the extended brightness depression at the center of the image.

Figure~\ref{fig:cross_sec} shows, as an illustrative case, the average image of a non-spinning MAD model and the cross-section of the image that is perpendicular to the spin axis of the black hole. In this example, applying our definition of the image brightness asymmetry yields the ratio of the integral of the blue shaded region to the integral of the red shaded region. Because this particular model is for a non-spinning black hole, the center of the image is at $X=0$. 

As a demonstration that the brightness asymmetry is caused by Doppler effects and, therefore, is maximized along the $Y=0$ cross section, Figure~\ref{fig:asym_angle} shows the magnitude of asymmetry $A$ but calculated along different cross-sections at various position angles with respect to the $X-$axis. In other words, a position angle of $90^{\circ}$ corresponds to a cross section that is perpendicular to the black hole spin axis. The various curves correspond to the mean image of all four GRMHD models at different inclination angles. For all simulations that have an inclination $\ge 0^{\circ}$, the maximum brightness asymmetry indeed occurs for position angles  $\simeq 90^{\circ}$.

The non-spinning MAD model is an exception to the above argument, as the mean images in this simulation have a peak asymmetry that is slightly offset from $90^{\circ}$ (see e.g., Figure~\ref{fig:ne_i0MAD}). Coherent asymmetries in the flow can arise from the buoyancy of the magnetic fields in the disk and can remain stable over several dynamical timescales at the large radii that feed the inner accretion flow. Even though this simulation was run for a long time span, there is some persistent asymmetry in the flow parameters (e.g., $n_e$, $B$, $\beta$) above and below the $\theta=\pi/2$ plane, which results in the slight offset seen in Figure~\ref{fig:ne_i0MAD}.

%==============================================================================
\section{correlations between image brightness asymmetry and model parameters}\label{sec:results}

%%%%%%%%%%%%%%%%%%%%%%%%%%%%%%%%%
\begin{figure}[t!]
\centering
\includegraphics[width=\columnwidth]{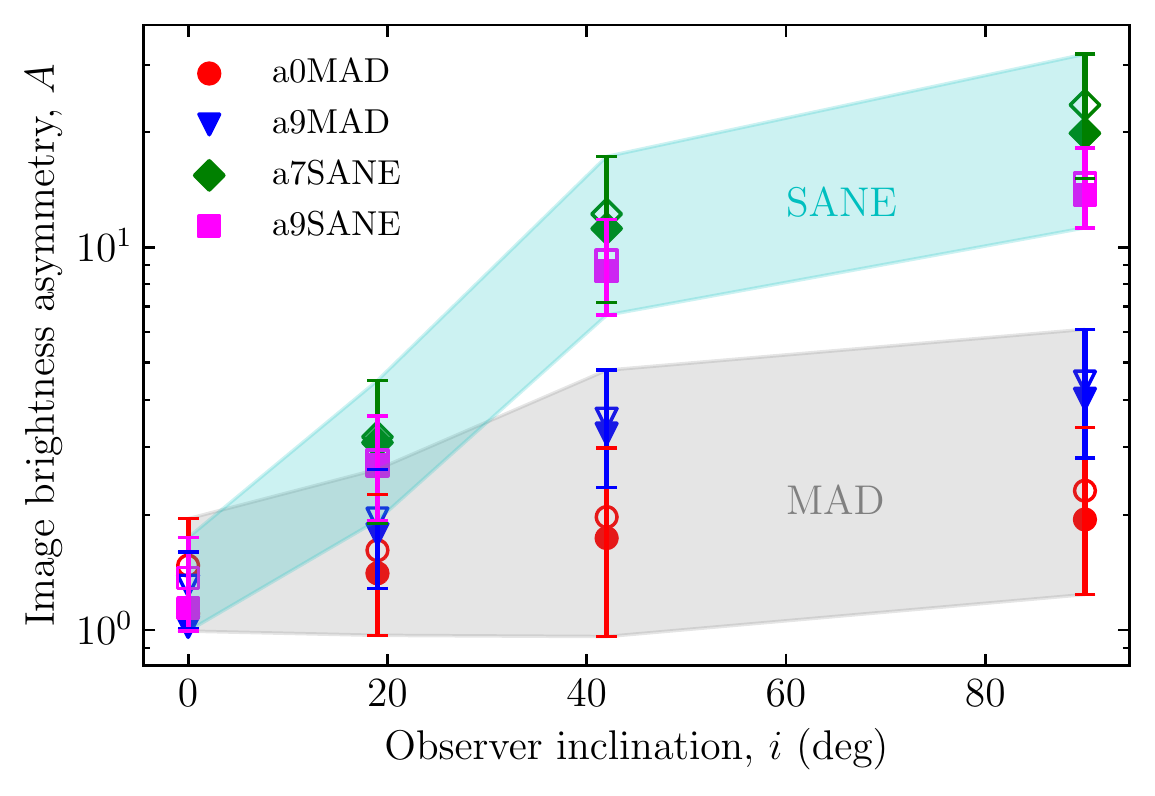}
\caption{Image brightness asymmetry as a function of observer inclination angle for all simulations in our set with $n_{e,0}=10^7\mathrm{cm}^{-3}$ and $R_{\mathrm{high}}=20$. Here and in the following figures, filled markers correspond to the image brightness asymmetry of the mean image of that simulation and empty markers and error bars correspond to the mean and standard deviation of the distribution of image brightness asymmetry calculated for each snapshot in the simulations. All models show increased brightness asymmetry with inclination, as expected for Doppler effects. The degree of asymmetry at high inclinations, however, is markedly different between the SANE (cyan shaded region) and the MAD (gray shaded region) models.}
\label{fig:asym_i}
\end{figure}

In order to explore the dependence of image brightness asymmetry on the various parameters of the models and of the black hole, we have calculated it for all of the individual 1024 snapshots of each simulation as well as for all mean images of each simulation. Figure \ref{fig:asym_i} shows the image brightness asymmetry as a function of the inclination angle of the observer, both for the mean images and for the individual snapshots. Even though there is some appreciable variance in the asymmetry between snapshots of the same simulation caused by the turbulent nature of the flow, the difference between MAD and SANE models as well as the dependence of the brightness asymmetry on the observer inclination introduce substantially larger variations. Indeed, the asymmetry in the mean images of each simulation provides an accurate measure of the typical asymmetry found in the individual snapshots.

\begin{figure}[t!]
\centering
\includegraphics[width=\columnwidth]{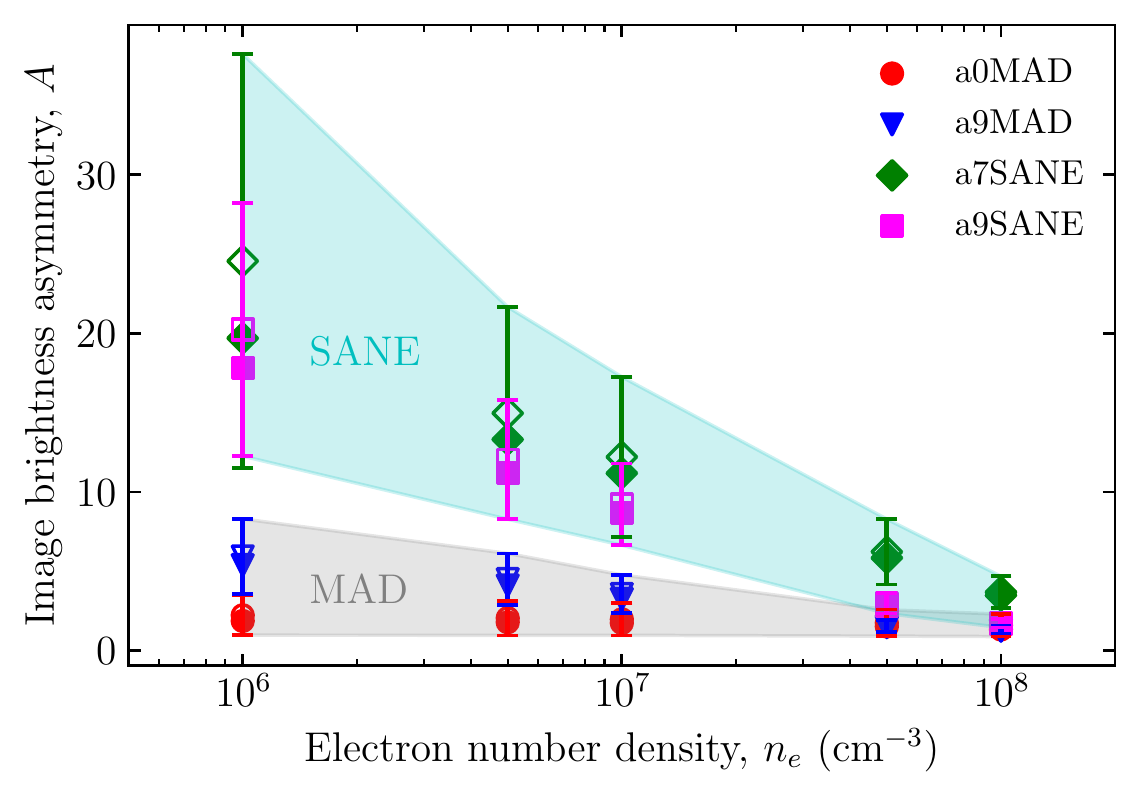}
\caption{Image brightness asymmetry as a function of the electron number density scale $n_{e,0}$ for all simulations in our set with $i=42^{\circ}$ and $R_{\mathrm{high}}=20$.}
\label{fig:asym_ne}
\end{figure}

\begin{figure}[t!]
\centering
\includegraphics[width=\columnwidth]{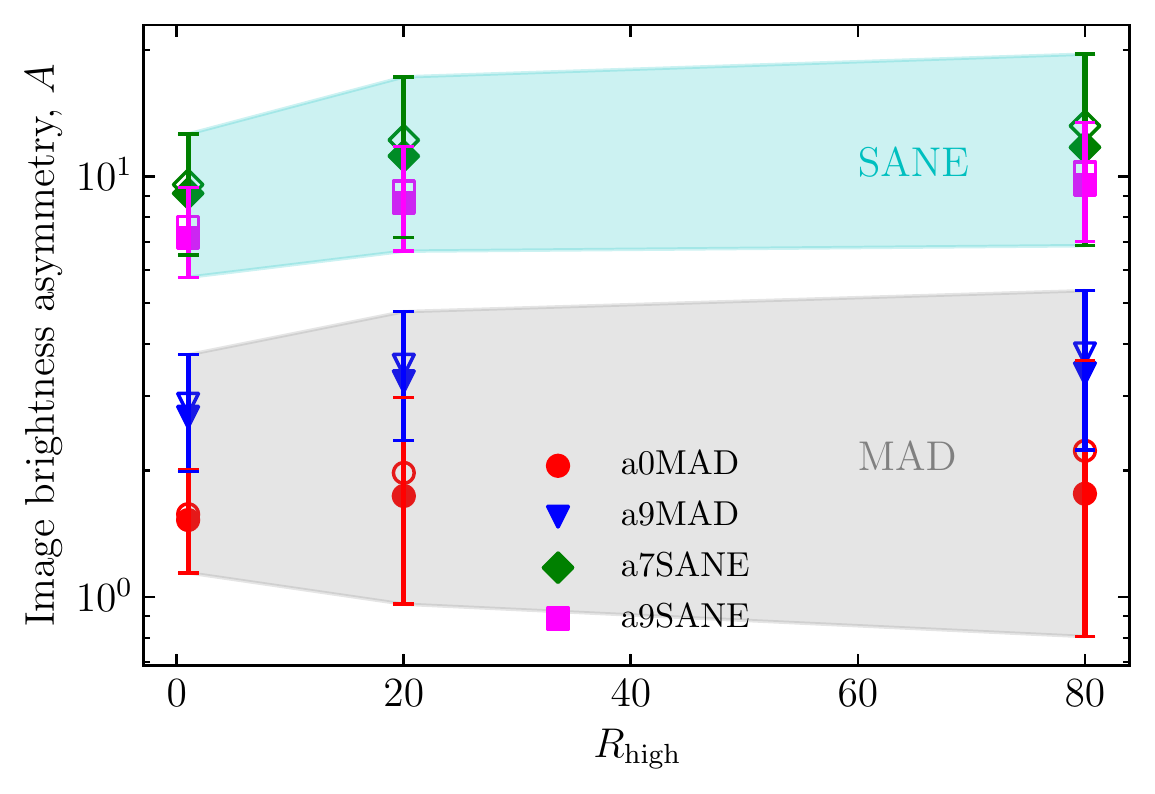}
\caption{Image brightness asymmetry as a function of $R_{\mathrm{high}}$ for all simulations in our set with $n_e=10^7$ and $i=42^{\circ}$.}
\label{fig:asym_R}
\end{figure}

At each inclination, SANE models have consistently higher asymmetry than the two MAD models, as expected by the fact that the plasma velocities in the former are significantly larger than in the latter (see also Figures \ref{fig:asym_ne} and \ref{fig:asym_R}). For the same reason, the brightness asymmetry of the SANE models increases significantly with observer inclination, whereas the MAD models maintain relatively low asymmetry even at high inclinations. In fact, the non-spinning MAD model viewed edge on has an asymmetry that is $\sim 10$ times lower than the SANE models viewed at the same inclination. This figure serves as a demonstration of the fact that an image with low brightness asymmetry does not require low observer inclination. On the other hand, if high asymmetry is observed over several epochs, it will be indicative of not only a high inclination but also of the presence of large azimuthal plasma velocities.
%%%%%%%%%%%%%%%%%%%%%%%%%%%%%%%%%

Figure~\ref{fig:asym_ne} shows the dependence of image brightness asymmetry on the electron number density scale $n_{e,0}$. As discussed earlier, when the electron number density increases in the flow, so does the width of the ring in the image. Thicker rings result in more symmetric images since the Doppler effects and lensing effects are less dominant in determining the image structure. It is interesting that the variance in the brightness asymmetry between the snapshots of the various simulations increases with decreasing value of the electron number density scale, especially for the SANE models. This is expected because, when the ring width in the image is small, localized turbulent perturbations in the plasma emission from regions with high velocities cause substantial brightness changes in the image. On the other hand. when the ring width in the image is large and the emission is coming also from regions of smaller plasma velocities, the effect of several localized perturbations is averaged out and the brightness asymmetry becomes less variable.

Finally, Figure \ref{fig:asym_R} shows the dependence of image brightness asymmetry on the ion-to-electron temperature ratio $R_{\mathrm{high}}$. As discussed earlier, changing $R_{\mathrm{high}}$ has a significantly smaller effect on image brightness asymmetry than the other parameters, which is within the variance seen for each model between the different image snapshots.

%%%%%%%%%%%%%%%%%%%%%%%%%%%%%%%%%

%%%%%%%%%%%%%%%%%%%%%%%%%%%%%%%%%
\begin{figure}[t!]
\centering
\includegraphics[width=\columnwidth]{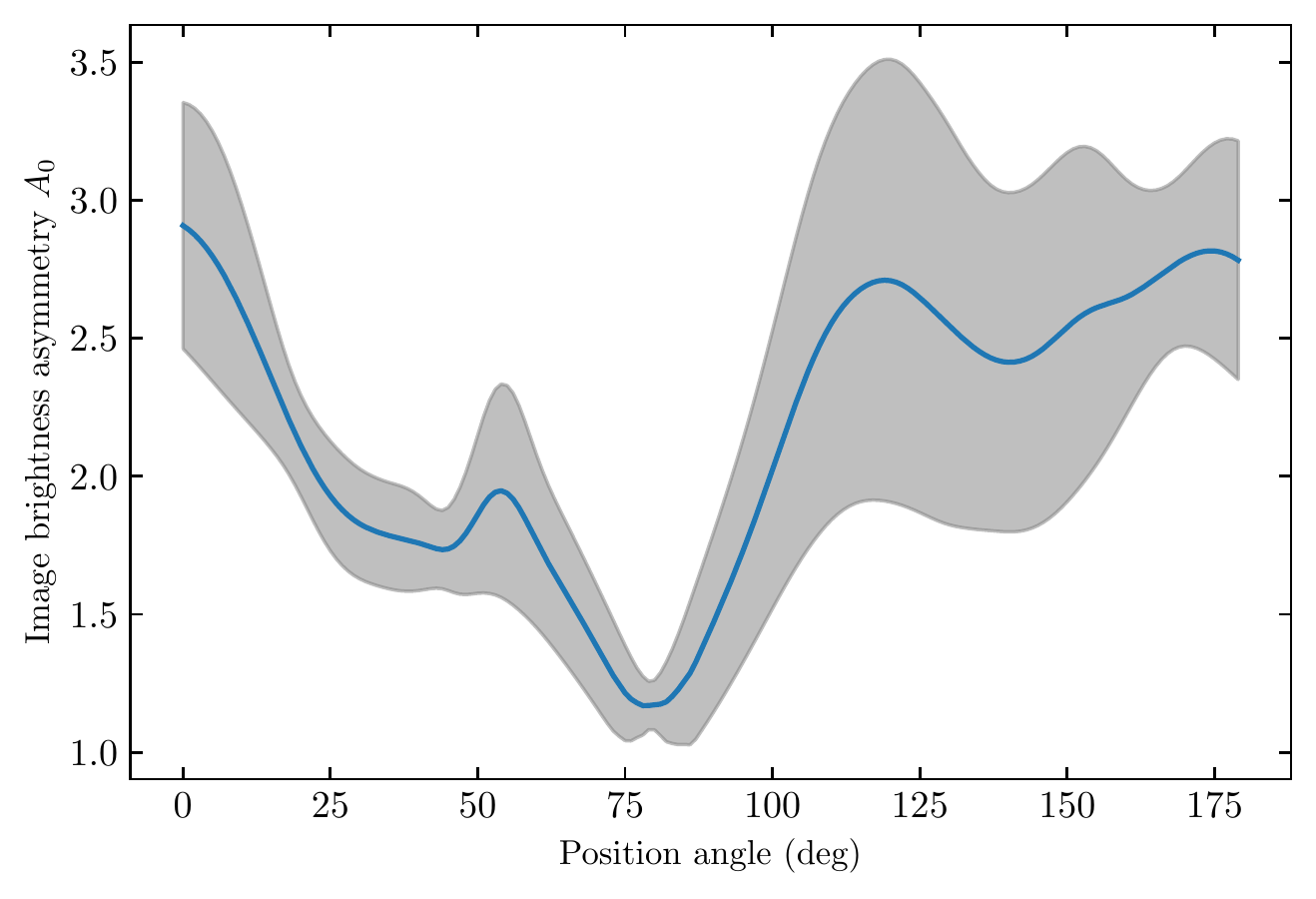}
\caption{Mean image brightness asymmetry (blue curve) for a set of EHT M87 images as a function of the position angle of the cross section used to measure asymmetry. The 68-percentile contours are shown in gray. We define position angle as the angle East of North such that $0^{\circ}$ corresponds to a North-South cross section. Note that there is a degeneracy between cross sections with position angles $\theta$ and $\theta+180^{\circ}$ because the asymmetry is defined to be above unity.}
\label{fig:asym_PA_M87}
\end{figure}

%==============================================================================
\section{Implications for M87}\label{sec:discussion}

We now consider the implications of our work for the images of M87 recently obtained by the EHT \citep{2019ApJ...875L...1E}. 
Because the EHT is a sparse interferometer, converting the interferometric data to images requires specialized algorithms with nuisance parameters, such as regularizers, that are tuned to the particular characteristics of the interferometer and the source (for details, see, \citealt{2019ApJ...875L...4E}). A number of such images have been generated, with three different imaging algorithms, while exploring a wide range of values for the nuisance parameters. All images are consistent with a narrow ring of emission with properties that depend very little on the particular details of image reconstruction and possess a small North-South brightness asymmetry \citep{2019ApJ...875L...4E}. 

In the previous sections, we showed that the maximum degree of asymmetry in a black-hole image depends on the inclination of the black hole spin with respect to the observer's line of sight. In M87, there is an {\it a priori} inference of the orientation of the black-hole spin based on the properties of the large scale jet. The position angle in the sky of the jet has been estimated to be $288^{\circ}$ East of North \citep{2018ApJ...855..128W}, while its inclination has been inferred to be $17^{\circ}$ with respect to our line of sight \citep{2018ApJ...855..128W}.  It is likely that the spin axis of the black hole is aligned with the large scale jet (see, however, \citealt{2020MNRAS.499..362C} for a discussion of jet alignment in the case of tilted disks). In this section, we compare this information to the asymmetry properties of the EHT images. Our aim is not to show that a different inclination angle is more likely for the case of M87 but rather ask whether low-inclination is required in general to explain the low asymmetry seen in the images. 

In order to compare the results of our simulations to the outcome of the observations, we apply our image brightness asymmetry measure to a fiducial set of reconstructed EHT images. Figure~\ref{fig:asym_PA_M87} shows the dependence of the brightness asymmetry on the position angle of the cross section. Our analysis did not include an in-depth exploration of the effect of image reconstruction of sparse interferometric data on image brightness asymmetry. Therefore, we do not focus on the details of this dependence but limit ourselves to only a qualitative comparison with theoretical expectations. The mean asymmetry for the fiducial set of EHT images for a large subset of position angles is above $A=2$ and the maximum asymmetry is $A=2.9$; we choose $A=2.5$ as a representative value. 

We ran a new, focused set of simulations with parameters that are appropriate for M87. Specifically, we set the black hole mass to $M_{\mathrm{BH}}=6.5\times10^9M_{\odot}$, added an intermediate inclination value of $i=17^{\circ}$, and probed electron number density scales in the range $n_{e,0}=10^5 \, \,\mathrm{to}\,\,10^6\mathrm{cm}^{-3}$ (see \citealt{2015ApJ...799....1C} for a description of these parameters). For this new set of tailored simulations, we also applied a Butterworth filter to the images, which suppresses power above $8\mathrm{G}\lambda$, the length of the longest EHT baseline (see  \citealt{2020arXiv200406210P} for details on this filter). This allows a more direct comparison with the observed images. 

%%%%%%%%%%%%%%%%%%%%%%%%%%%%%%%%%
\begin{figure}[t!]
\centering
\includegraphics[width=\columnwidth]{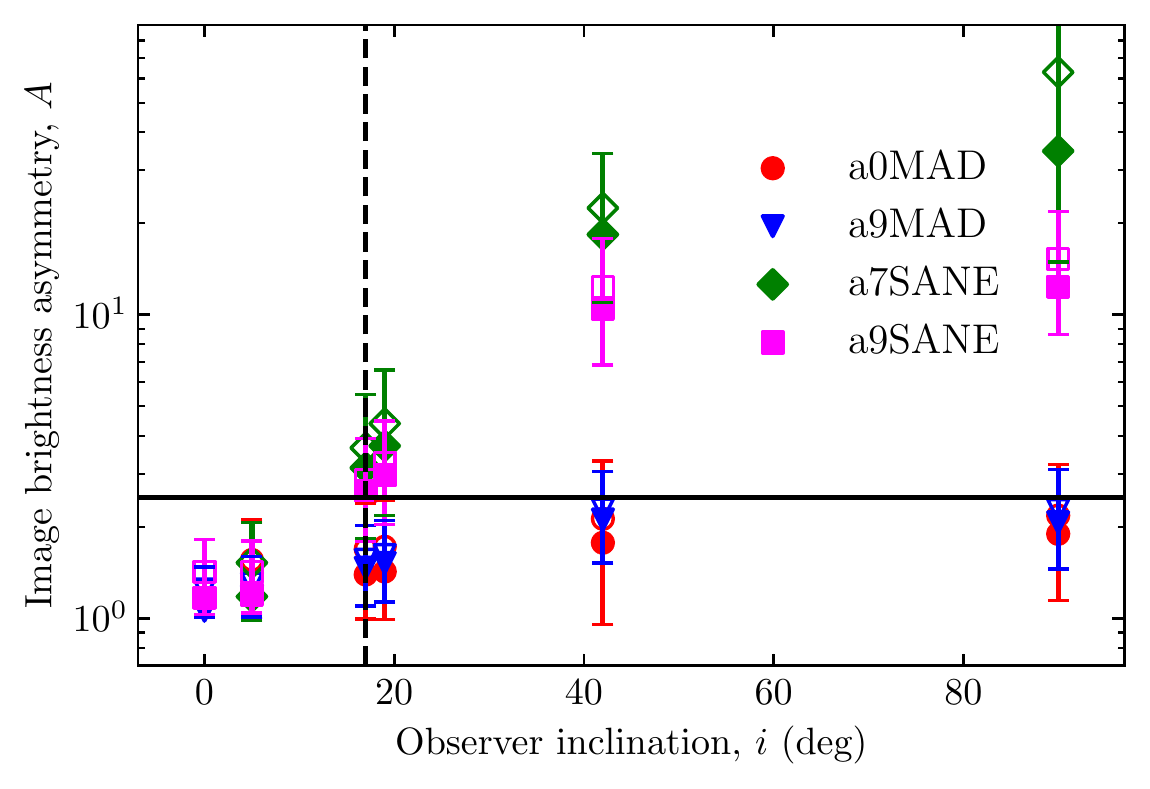}
\caption{Image brightness asymmetry as a function of inclination angle $i$ for simulations with $R_{\mathrm{high}}=20$, $n_{e,0}=5\times10^5\mathrm{cm}^{-3}$, and a black-hole mass of  $M_{\mathrm{BH}}=6.5\times10^9M_{\odot}$. The simulation images have been filtered with a Butterworth filter that removes most power above 8$\mathrm{G}\lambda$ before calculating the asymmetry ratio for a better comparison with EHT images. The black line at brightness asymmetry of 2.5 shows the approximate asymmetry in the set of M87 images reconstructed from the 2017 EHT observations. The vertical dashed line corresponds to the $\sim 17^\circ$ degree inclination of the M87 jet. Even though both SANE and MAD models are consistent with the observed brightness asymmetry at this inclination, MAD models would have been able to reproduce such a low asymmetry even when viewed edge on.}
\label{fig:asym_iM87}
\end{figure}

In Figure~\ref{fig:asym_iM87}, we plot the image brightness asymmetry as a function of the observer inclination obtained from this simulation library and compare this to the representative value of the asymmetry in the M87 images. It is clear from this figure that, even though both SANE and MAD models are consistent with the observed brightness asymmetry at the inferred $17^\circ$ inclination of the M87 black hole, MAD models would have been able to reproduce such a low asymmetry even when viewed edge on. This argues against using the image symmetry as a direct probe of observer inclination with respect to the black-hole spin.

%==============================================================================
\section{Summary}\label{sec:summary}

The images of black holes generated by the EHT have a number of coarse-scale properties that can lead to general inferences about the black holes and their accretion flows that are only marginally model dependent. For example, the size of the bright emission ring has been used to infer the mass of the black hole and test the predictions of the theory of General Relativity~\citep{2019ApJ...875L...6E,2020PhRvL.125n1104P}. In the same spirit, the presence of a brightness asymmetry around the emission ring has been used in the past as an indication of the observer inclination with respect to the angular momentum of the accretion flow and, perhaps, of the black hole itself; it has been applied both to early observations of Sgr~A$^*$ \citep{2015ApJ...798...15P} as well as to the most recent images of the black hole in the center of M87 \citep{2019ApJ...875L...5E}.

In this paper, we explored how image brightness asymmetry is related to various black hole and model parameters and reevaluate the early argument that images with low asymmetry can only be generated in sources viewed at low inclinations, i.e., nearly face on. We use a large suite of GRMHD simulations to find explicit counter examples to this argument, in which images with low brightness asymmetry are generated even for high observer inclinations in models that have accumulated substantial magnetic flux close to the black hole horizon, i.e., the so-called Magnetically Arrested Disks. The weak brightness asymmetry in these models is a consequence of the fact that the magnetic stresses significantly decrease the angular velocity of the plasma in the inner accretion flow and, therefore, also the effects of relativistic Doppler beaming and boosting. In particular, we show that low-spin MAD models have the lowest brightness asymmetry, generating nearly uniform ring images surrounding the black hole shadows.

%==============================================================================
\acknowledgements
L.\;M.\ gratefully acknowledges support from an NSF Astronomy and Astrophysics Postdoctoral Fellowship under award no. AST-1903847.
D.\;P.\,,  F.\;O.\,, and R.\;N.\, gratefully acknowledge support from NSF PIRE 
grant OISE-1743747, NSF AST-1715061. All ray tracing calculations were performed with the \texttt{El~Gato}
GPU cluster at the University of Arizona that is funded by NSF award
1228509.

\bibliographystyle{aasjournal}
\bibliography{main}
\end{document}